\documentclass[pra,twocolumn,showkeys,a4paper,showpacs,superscriptaddress]{revtex4}
\usepackage{amsmath}
\usepackage{amsfonts}
\usepackage{graphicx}
\begin{document}
\title{Local invariants and pairwise entanglement in symmetric multiqubit system}
\author{A. R. Usha Devi}
 \email{arutth@rediffmail.com}
\author{ M. S. Uma}
\author{ R. Prabhu}
\affiliation{
Department of Physics, Bangalore University, Bangalore-560 056, India\\}
\author{Sudha}
\affiliation{Department of Physics, Thunga First Grade College, \\ Thirthahalli-577432, India.}

\date{\today}
\begin{abstract}
Pairwise entanglement properties of a symmetric multi-qubit system
are analyzed through a complete set of two-qubit local
invariants. Collective features of entanglement, such as {\em spin squeezing},
are expressed in terms of invariants and a classification scheme for pairwise
entanglement is proposed. The invariant criteria given here are shown to be related to the 
recently proposed\,({\it Phys. Rev. Lett.} {\bf 95}, 120502 (2005))\,generalized spin squeezing 
inequalities for pairwise entanglement in symmetric multi-qubit states.
\end{abstract}

\maketitle

\section{\label{sec:level1}Introduction}
Considerable interest has been evinced
recently~\cite{Wineland92,Kuz97,Kuz98,Pol99,Hald99,Kuz00,Duan00,Koz00,Sor101,Ficek02,Duan03} in
producing, controlling and manipulating entangled multi-atom systems due to the possibility of 
applications in atomic
interferometry~\cite{Yur86,Kit91}, high precession
atomic clocks~\cite{Wineland94}, quantum computation and quantum
information processing~\cite{Nie00}. Multi-atom systems, which are
symmetric under permutation of the particles, allow for an elegant
description in terms of collective variables of the system.
Specifically, if we have $N$ two level atoms, each atom may be
represented as a spin-$\frac{1}{2}$ system (a qubit) and
theoretical analysis in terms of collective spin operator
$\vec{S}=\frac{1}{2}\displaystyle\sum_{\alpha=1}^N\ \vec{\sigma}_\alpha ,$ \ \
($\vec{\sigma}_\alpha$ denote the Pauli spin operator of the $\alpha^{\rm th}$
qubit),
leads to  reduction of the dimension of the Hilbert space from $2^N$ to
$(N+1)$, when
the two level multi-atom system respects exchange symmetry. A large number of
experimentally
relevant multi-atom states exhibit symmetry under interchange of atoms,
facilitating a significant simplification in understanding the properties of
the
physical system.

While complete characterization of multi-particle entanglement still remains a major task, 
collective behaviour such as  {\em spin 
squeezing}~\cite{Wineland92,Kuz97,Kuz98,Pol99,Hald99,Kuz00,Duan00,Koz00,Sor101,Kit93,Agarwal90,
Lukin00,Xwang01,Sor201,Orzel01,Sor02,Andre02,Usha103,Usha203,Ulam01,Xwang03}, exhibited by 
multi-atom systems, has been proposed as a signature of quantum correlation
between the atoms. A connection between spin squeezing and the nature of
quantum entanglement has been explored~\cite{Ulam01,Xwang03} and it is shown
that the presence of spin squeezing essentially reflects pairwise entanglement. However,
it is important to realize that spin squeezing serves only as a sufficient condition - not a 
necessary one - for pairwise entanglement. There will
still be pairwise correlated states, which do not exhibit spin squeezing.  In a
class of symmetric multiqubit states it has been shown~\cite{Xwang03}  that spin-squeezing and 
pairwise entanglement imply each other. Questions like
``{\em Are there any other collective  signatures of pairwise entanglement?}\,"
are still being investigated. Recently, inequalities generalizing the concept of spin squeezing 
have been derived~\cite{cirac}. These inequalities are shown to provide necessary and sufficient 
conditions for pairwise entanglement and three-party entanglement in symmetric $N$-qubit states. 
In this paper, we propose a complete characterization and classification of pairwise entanglement 
in symmetric multi-qubit systems, in terms
of local invariants associated with a random pair of qubits drawn from the
collective system. We show that
pairwise entanglement in symmetric  $N$-qubit states is reflected through {\em
negative} values of some of the
two-qubit invariants. Specifically, a symmetric multi-qubit system is spin
squeezed {\em iff}\,  one of the entanglement
invariant is {\em negative}. We discuss a classification scheme for pairwise entanglement in 
symmetric $N$-qubit states based on {\em non-positive} values of the local invariants. We show 
that our characterization is indeed related to the generalized spin squeezing inequalities of 
Ref.~\cite{cirac}. In the light of our characterization, we analyze physical examples of 
symmetric
multi-qubit states like (i) Dicke state,
(ii) Kitagawa-Ueda state generated by one axis twisting Hamiltonian and (iii)
Steady state of atoms irradiated
by a squeezed bath.

\section{\label{sec:level1} Local Invariants of a Symmetric Two-qubit system}
A symmetric $N$-qubit system remains unchanged by permutations of the qubits:
$$\Pi_{\alpha\beta}\, \rho_N\, \Pi_{\alpha\beta}^{-1}=\rho_N, $$
where $\Pi_{\alpha\beta}$ denotes the permutation  operator interchanging
$\alpha^{\rm th}$ and $\beta^{\rm th}$ qubits
of the multi-qubit system. Owing to this symmetry, quantum states of symmetric
multi-qubits get restricted to a $(2S+1)$
dimensional subspace $\left\{\ \vert S=\frac{N}{2},\ M\rangle ;\, -S\leq M\leq
S\right\}$  of the entire
$2^N$ dimensional Hilbert space $C^2\otimes C^2\ldots \otimes C^2$~.
Here, $\vert S,\ M\rangle$ are the
simultaneous eigen states of the squared collective spin operator $S^2$ and
$S_3$, the $z$-component of the collective operator
$\vec{S}=\frac{1}{2}\, \displaystyle\sum_{\alpha=1}^N \vec{\sigma}_\alpha$.

In a symmetric $N$-qubit system the average values of collective spin
correlations, upto second order, are expressed in terms of their
two-qubit counterparts as follows:
\begin{equation}
\label{vars}
\langle S_i\rangle=\frac{1}{2}\sum_{\alpha=1}^N \langle \sigma_{\alpha
i}\rangle  =\frac{N}{2}s_i
\end{equation}
and
\begin{eqnarray}
\label{vart}
\frac{1}{2}\langle (S_iS_j+S_jS_i)\rangle&=& \frac{1}{8}\,
\displaystyle\sum_{\alpha,\beta=1}^N\left\langle (\sigma_{\alpha
i}\sigma_{\beta j}
+\sigma_{\beta i}\sigma_{\alpha j})\right\rangle \nonumber\\
&=&\frac{1}{4}\displaystyle
\sum_{\alpha,\beta=1}^N\left\langle (\sigma_{\alpha i}\sigma_{\beta
j})\right\rangle \nonumber \\
&=&\frac{N}{4}\, \delta_{i\, j}+\frac{N(N-1)}{4}\left\langle (\sigma_{1
i}\sigma_{2 j})\right\rangle \nonumber \\
&=&\frac{N}{4}\,\left[ \delta_{i\, j}+ (N-1)\, t_{ij}\, \right ],\,\,\,\, 
i,j=1,2,3.\nonumber\\
& &
\end{eqnarray}
In the above equation, we have denoted the expectation values of the two qubit
observables by~\cite{footnote1}
 $$\langle\sigma_{1\,i}\rangle=\langle\sigma_{2\,
i}\rangle=s_i,\ \ \   {\rm and}$$
$$\langle(\sigma_{1\,i}\sigma_{2\,
j})\rangle=\langle(\sigma_{1j}\sigma_{2i})\rangle=t_{ij};\,\,\,\, 
i,j=1,2,3.$$
Here, $s_{i}$ are  the components of average spins of the qubits and
$t_{ij}=t_{ji}$ are the  elements of
the $3\times 3$ real, symmetric two-qubit correlation matrix.
Pairwise entanglement properties of a symmetric multiqubit state are
characterised in terms of the    symmetric two-qubit density matrix given by
\begin{equation}
 \label{rho}
\rho^{\rm sym}=\frac{1}{4}\left( 1+\displaystyle\sum_{i=1}^3
(\sigma_{1i}+\sigma_{2i})\, s_i+
\displaystyle\sum_{i,j=1}^3 (\sigma_{1i}\sigma_{2j})\, t_{ij}\right),
\end{equation}
which is associated with any random pair of qubits  belonging to a symmetric
multi-qubit ensemble.

Since the squared collective spin operator should satisfy the condition,
 $\displaystyle\sum_{i=1}^3~\langle S_i^2\rangle = S(S+1)=\frac{N}{2}
\left(\frac{N}{2}+1\right)$ we have
$\frac{1}{4}\langle(\vec{\sigma}_1+\vec{\sigma}_2)^2\rangle = 2$ at the
two-qubit level, which
leads to the constraint ${\rm Tr}(T)=1$, on the two-qubit correlation matrix
$T$. Therefore,
8 state parameters - viz., three components $s_i$ of the average qubit
orientation
and the elements $t_{ij}$ of the real, symmetric correlation matrix with unit
trace
- characterize the two-qubit partition of the system completely.

Entanglement of a composite quantum system remain invariant, when the
subsystems are subjected to local
unitary operations. Any two quantum states are entanglementwise equivalent iff
they are related to
each other through local unitary transformations.
Non-separability of a quantum state can be represented  by  complete set of
local invariants~\cite{Linden99,Sudbery01,Makhlin03,Ser03} 
and any quantitative  measure of
entanglement
must be a function of these invariants.
In the following Theorem, we show that a set containing  six local invariants
provide a complete characterization of entanglement in a symmetric two-qubit
system. This result proves to be
useful in analysing pairwise entanglement properties of an arbitrary symmetric
multi-qubit system.

\noindent{\bf Theorem 1.}
\label{iden}
{\em All equally entangled symmetric two-qubit states have identical values for the
local invariants
$\{ {\cal I}_1 \, -\, {\cal I}_6 \}$ given below:
\begin{eqnarray}
\label{inv}
&{\cal I}_1={\rm det}\, T\, ,\ \     {\cal I}_2={\rm Tr}\, (T^2)\, ,& \nonumber
\\
& {\cal I}_3=s^T\, s\, , \ \  \    {\cal I}_4=s^T\, T\, s\, , \nonumber \\
&{\cal I}_{5} = \epsilon_{ijk}\,\epsilon_{lmn}\, s_i\,s_l\, t_{jm}\, t_{kn} \,
, \nonumber \\
& {\cal I}_{6}=\epsilon_{ijk}\, s_i\, (T\, s)_j\, (T^2\,s)_k \, ,
\end{eqnarray}
where $\epsilon_{ijk}$ denotes Levi-Civita symbol; $s\ (s^T)$ is a column (row)
with
$s_1,\, s_2$ and $s_3$ as elements.}

\noindent{\bf Proof:}
Let us first note that the state parameters of a symmetric two-qubit density
matrix transform under
{\em identical local unitary operation}  ~\cite{footnote2} $U\otimes U$ as follows:
\begin{eqnarray}
\label{tran}
&s'_i=\displaystyle\sum_{j=1}^3 O^{(1)}_{ij}\, s_j   \hskip 0.1in   {\rm \ or \ \ }  s'=O\, s \,
, \hskip 0.9in    \nonumber \\
&t'_{ij}=\displaystyle\sum_{k,l=1}^3\, O^{(1)}_{ik}\, O_{jl}^{(2)}\, t_{kl} \hskip 0.2in   {\rm
or \ } \ \      T'=O\, T\,  O^T\, ,
    \end{eqnarray}
where $O\in SO(3, R)$ denotes $3\times 3$ rotation matrix, corresponding
uniquely to the $2\times 2$ unitary matrix
$U\in SU(2)$.

It is easy to verify that   $\{ {\cal I}_1 \, -\, {\cal I}_6 \}$   given by
~Eq.~(\ref{inv}) are all invariant, when the
state variables $s$ and $T$ transform under {\it identical local rotations} as
shown in ~Eq.~(\ref{tran}).
We may choose to specify the state parameters $s$ and $T$ in a basis in which
$T$ is diagonal. This is possible because
the real, symmetric correlation matrix $T$ can be diagonalized through
identical local rotations:
\begin{equation}
\label{diagT}
T^d=O\, T\, O^T={\rm diag}\, (t_1,\, t_2,\, t_3).
\end{equation}
It is clear that the invariants ${\cal I}_1,\ {\cal I}_2,$ (along with the unit
trace condition Tr\,$(T)$=1), determine the eigenvalues
$t_1,\ t_2$ and $t_3$ of the two-qubit correlation matrix $T$. Further, the
absolute values of the state variables $s_1,\ s_2,\ s_3,$ can be
evaluated using ${\cal I}_3,\ {\cal I}_4$ and ${\cal I}_5$:
\begin{eqnarray}
\label{s1s2s3}
{\cal I}_3&=&s_1^2+s_2^2+s_3^2, \nonumber \\
{\cal I}_4&=&s_1^2\, t_1+s_2^2\, t_2+s_3^2\, t_3, \nonumber \\
{\cal I}_5&=& 2\,(s_1^2\, t_2\,t_3+s_2^2\, t_1\, t_3+s_3^2\, t_1\, t_2) .
\end{eqnarray}
The overall sign of the product $s_1 s_2 s_3$ is then fixed by ${\cal I}_6$:
\begin{equation}
\label{sign}
{\cal I}_6=s_1 s_2 s_3\, \left[t_1\, t_2\, (t_2-t_1)+t_2\, t_3\,
(t_3-t_2)+t_3\, t_1\, (t_1-t_3)\right].
\end{equation}
It is important to realize that only the overall sign of $s_1\, s_2\, s_3$ -
not the individual signs - is a local invariant.
More explicitly, if $(+,+,+)$ denote the signs of $s_1,\, s_2\, {\rm \ and \ }
s_3,$ identical local rotation through an angle
$\pi$ about the axes $x,\, y\, {\rm \ or}\ z$ affect only the signs, not the
magnitudes of $s_1,\, s_2,\,  s_3,$ leading to the
possibilities  $(+,-,-)$, $(-,+,-)$, $(-,-,+).$ All these combinations
correspond to the `$+$' sign for the product $s_1\, s_2\, s_3$.
Similarly, the overall `$-$' sign for  the product  $s_1 s_2 s_3$ arises from the
combinations, $(-,-,-)$, $(-,+,+)$, $(+,-,+),$
$(+, +, -)$, which are all related to each other by $180^{\circ}$ local
rotations about the  $x,\ y {\rm\  or}\ z$ axes. 

Thus we have shown that every symmetric two qubit density matrix can be
transformed by identical local unitary transformation
$U\otimes U$ to a {\em canonical form}, specified completely by the set of
invariants $\{ {\cal I}_1\, -\, {\cal I}_6\}$. In other words,
symmetric two-qubit states are equally entangled iff $\{ {\cal I}_1\, -\, {\cal I}_6\}$ are 
same.$\Box$

Now, we proceed to  identify  criteria of pairwise entanglement in a symmetric
multi-qubit state, in terms of the two-qubit
local invariants  $\{ {\cal I}_1\, -\, {\cal I}_6\}$.

We first note that the invariants ${\cal I}_2,\ {\cal I}_3 \geq 0$ for all
symmetric states. The signs of other invariants play a
significant role in characterizing entanglement.\\
\noindent{\bf Theorem 2.}
\label{nonneg}
{\em The invariants ${\cal I}_1,\ {\cal I}_4,\ {\cal I}_5$ and the combination
${\cal I}_4-{\cal I}_3^2$ are {\em non-negative} in
a symmetric separable state.}

\noindent{\bf Proof:}
A separable symmetric multi-qubit state is given by
\begin{equation}
\label{sep}
\rho_{\rm (sym-sep)}= \displaystyle\sum_w\, p_w\, \rho^{(w)}\otimes
\rho^{(w)}\otimes\ldots \rho^{(w)},
\end{equation}
with $\displaystyle\sum_w p_w =1$ and
$\rho^{(w)}=\frac{1}{2}\, \left(1+\displaystyle\sum_{i=1}^3\, \sigma_i\ s_i^w \right)$
denote single qubit density matrix. Two qubit partition of the separable
symmetric system is given by
\begin{eqnarray}
\label{sep2}
\rho_{\rm (sym-sep)}^{(\alpha\beta)}&=&{\rm Tr}_{\{1,2,\ldots, N /\alpha,
\beta\} }\, (\rho_{\rm (sym-sep)})\nonumber \\
&=&\displaystyle\sum_w p_w \, \rho^{(w)}\otimes \rho^{(w)}.
\end{eqnarray}
In ~Eq.~(\ref{sep2}), the two-qubit density matrix $\rho_{\rm
(sym-sep)}^{(\alpha\beta)}$ is independent of the qubit indices $\alpha,\
\beta$, owing to
exchange symmetry. The state variables of the two-qubit separable symmetric
state are given by
\begin{eqnarray}
\label{sepst}
s_i&=&{\rm Tr}\left(\rho_{\rm (sym-sep)}^{(\alpha\beta)}\, \sigma_i\right)
=\displaystyle\sum_w p_w\, s_i^{(w)},\\
t_{ij}&=& {\rm Tr}\left(\rho_{\rm (sym-sep)}^{(\alpha\beta)}\, \sigma_{\alpha
i} \sigma_{\beta j}\right)
=\displaystyle\sum_w p_w\, s_i^{(w)}\, s_j^{(w)}.
\end{eqnarray}

\noindent (i) To prove that ${\cal I}_1\geq 0$ in a separable symmetric state,
we transform $\rho_{\rm (sym-sep)}^{(\alpha\beta)}$
with the help of a suitable local rotation such that $T={\rm diag}(t_1,\,
t_2,\, t_3)=
{\rm diag}\left(\displaystyle\sum_w p_w\, \left(s_1^{(w)}\right)^2,\
\displaystyle\sum_w p_w\, \left(s_2^{(w)}\right)^2, \
\displaystyle\sum_w p_w\, \left(s_3^{(w)}\right)^2\right).$ We therefore have
\begin{equation}
\label{I1}
{\cal I}_1=\det T=t_1\, t_2\, t_3
=\displaystyle\prod_{i=1}^3\left(\displaystyle\sum_w p_w\,
\left(s_i^{(w)}\right)^2\right),
\end{equation}
which is obviously non-negative.

\noindent(ii)  The invariant ${\cal I}_4$ has the following structure for  a
separable state:
\begin{eqnarray}
\label{I4}
{\cal I}_4&=&s^T\, T\, s=\displaystyle\sum_{i,j=1}^3 t_{ij}\, s_i\,
s_j\nonumber \\
&=& \displaystyle\sum_w p_w\, \left(\displaystyle\sum_{i=1}^3 s_i^{(w)}\,
s_i\right)
         \left(\displaystyle\sum_{j=1}^3 s_j^{(w)}\, s_j\right) \nonumber \\
&=& \displaystyle\sum_w p_w\, \left(\vec{s}\cdot\vec{s}^{(w)}\right)^2 \geq 0.
\end{eqnarray}

\noindent(iii) Now, consider  the invariant ${\cal I}_5$ for a separable
symmetric system:
\begin{eqnarray}
\label{I5}
{\cal I}_{5} &=& \epsilon_{ijk}\,\epsilon_{lmn}\, s_i\,s_l\, t_{jm}\,
t_{kn}\nonumber \\
             &=&\displaystyle\sum_{w,w'} p_w\, p_{w'}\, \left(\epsilon_{ijk}\, s_i\,
s_j^{(w)}\, s_k^{(w')}\right)
               \left(\epsilon_{lmn}\, s_l\, s_m^{(w)}\, s_n^{(w')}\right)
\nonumber \\
             &=& \displaystyle\sum_{w,w'} p_w\, p_{w'}\, \left[ \vec{s}\cdot \left(
\vec{s}^{(w)}\times \vec{s}^{(w')}\right) \right]^2 \geq 0.
\end{eqnarray}

\noindent(iv) For the combination ${\cal I}_4-{\cal I}_3^2$ we obtain,
\begin{equation}
\label{comb}
{\cal I}_4-{\cal I}_3^2= \displaystyle\sum_w p_w\,
\left(\vec{s}\cdot\vec{s}^{(w)}\right)^2
               -\left(\displaystyle\sum_w p_w\,
(\vec{s}\cdot\vec{s}^{(w)})\right)^2,
\end{equation}
which has the structure $\langle A^2\rangle -\langle A\rangle ^2$ and is
therefore, essentially non-negative.$\Box$

Negative value assumed by any of the invariants ${\cal I}_1,\ {\cal I}_4,\
{\cal I}_5$ or the
${\cal I}_3-{\cal I}_4^2$, is a signature of pairwise entanglement. Moreover,
from  the structure of the invariants in  a symmetric separable state,
it is clear that  ${\cal I}_4=\displaystyle\sum_w p_w\,
\left(\vec{s}\cdot\vec{s}^{(w)}\right)^2=0$ implies  $\vec{s}^{(w)}\equiv 0$
for all `$w$', leading
in turn to ${\cal I}_3=\displaystyle\sum_w p_w\,
\left(\vec{s}\cdot\vec{s}^{(w)}\right)=0$ and
  ${\cal I}_5=0$ (see Eq.~(\ref{I5})).  So,  ${\cal I}_3\neq 0$,
${\cal I}_4, \ {\cal I}_5 \leq 0,$  reflects  pairwise entanglement in symmetric
multi-qubit states.

In the next section, we express collective properties, which exhibit pairwise
entanglement, in terms of two-qubit invariants.
\section{\label{sec:level2}Collective Signatures of Pairwise Entanglement}
Collective phenomena, reflecting pairwise entanglement of qubits, should be
expressible through local invariants. We begin with {\em spin squeezing} in $N$-qubit symmetric 
states.
Spin squeezing is a manifestation of quantum correlations, resulting in the
reduced quantum fluctuations in one
of the collective spin components, normal to the mean spin direction, below the standard quantum 
limit $N/4$ of spin coherent
states~\cite{Kit93}. A quantitative measure of this feature is
given by the spin squeezing parameter,
\begin{equation}
\label{xi}
\xi^2= \frac{4}{N}\, (\Delta\, S_{\perp})^2_{\rm min},
\end{equation}
where $S_{\perp}= \vec{S}\cdot \hat{n}_{\perp}$ is a perpendicular component of
collective spin;  $\hat{n}_\perp$ denotes
a unit vector in the plane orthogonal to the mean spin direction $\langle
\vec{S}\rangle$, and is chosen so that the variance
$(\Delta\, S_{\perp})^2$ is minimized. Symmetric multiqubit states with
$\xi^2<1$ are spin squeezed. We now show that

\noindent{\bf Theorem 3.}
\label{spin}
{\em For all spin squeezed states, the local invariant ${\cal I}_5$ is negative.}\\
\noindent{\bf Proof:}
It is useful to evaluate the invariant ${\cal I}_5$ (see Eq.~(\ref{inv})),
after subjecting the quantum state to a identical local rotation, which is designed to align the 
average spin vector
$\langle \vec{S}\rangle$ along the $z$-axis i.e., for the two-qubit partition
of the system, the state variable
$\vec{s}\equiv (0,\ 0,\ s_0)$ after this local operation. We now obtain,
\begin{eqnarray}
\label{ssI5}
{\cal I}_{5} &=& \epsilon_{3jk}\,\epsilon_{3mn}\, s_0^2 \, t_{jm}\,
t_{kn}\nonumber \\
             &=& 2\, s_0^2\, (t_{11}t_{22}-t_{12}^2)\nonumber \\
             &=& 2\, s_0^2\, \det T_{\perp},
\end{eqnarray}
 where  $T_{\perp}$ denotes the $2\times 2$ block of the correlation matrix in
the subspace  orthogonal to the qubit orientation direction i.e., z-axis.
 Now, we can still exploit the freedom of local rotations in the
$x-y$ plane, which leaves the average  spin $\vec{s}= (0,\ 0,\ s_0)$
unaffected. We use this to diagonalize
$T_{\perp}$:
\begin{equation}
\label{tperp}
T_{\perp}=\left(\begin{array}{ll}t_\perp^{(+)} & 0 \\ 0 &
t_{\perp}^{(-)}\end{array}\right) ,
\end{equation}
with
$t_{\perp}^{(\pm)}=\frac{1}{2}\left[(t_{11}+t_{22})\pm\sqrt{(t_{11}-t_{22}^2)+4
\, t_{12}^2}\right].$
We thus obtain,
\begin{equation}
\label{ss2I5}
{\cal I}_5= 2\ s_{0}^2\ t_{\perp}^{(+)}\, t_{\perp}^{(-)}.
\end{equation}
We now express the spin squeezing parameter $\xi^2$, given by Eq.~(\ref{xi}),
in terms of the two-qubit
state parameters:
\begin{eqnarray}
 \label{xi2}
\xi^2&=& \frac{4}{N}\, \left\langle (\vec{S}\cdot
\hat{n}_{\perp})^2\right\rangle _{\rm min} \nonumber \\
     &=& \frac{1}{N}\, \displaystyle\sum_{\alpha,\beta=1}^N \, \left\langle
(\vec{\sigma}_{\alpha}\cdot\hat{n}_{\perp})\,
         (\vec{ \sigma}_{\beta}\cdot\hat{n}_{\perp})\right\rangle_{\rm min}
\nonumber \\
     & =&  1+\frac{1}{N} \displaystyle\sum_{\alpha=1}^N
\displaystyle\sum_{\beta\neq\alpha=1}^N
           \left\langle (\vec{\sigma}_{\alpha}\cdot\hat{n}_{\perp})\,
           (\vec{\sigma}_{\beta}\cdot\hat{n}_{\perp})\right\rangle_{\rm
min}\nonumber \\
      &=&  1+  \frac{2}{N}\, \displaystyle\sum_{\alpha=1}^N
\displaystyle\sum_{\beta>\alpha=1}^N
            \left(\displaystyle\sum_{i,j=1}^3 \, \left\langle (\sigma_{\alpha\,
i}\sigma_{\beta\, j})\right\rangle
           n_{\perp i}\, n_{\perp j}\right)_{\rm min}. \nonumber \\
	& & 	   
\end{eqnarray}
For a symmetric system, $\left\langle \sigma_{\alpha\, i}\sigma_{\beta\,
j}\right\rangle =t_{ij}$, - independent
of the qubit indices $\alpha, \ \beta$ -  and we obtain,
\begin{eqnarray}
\label{xi3}
\xi^2&=& 1+(N-1)\, \left(\displaystyle\sum_{i,j=1}^3 \, t_{ij}\, n_{\perp i}\,
n_{\perp j}\right)_{\rm min} \nonumber \\
     &=& 1+(N-1)\, (n_{\perp}^T\, T\, n_{\perp})_{\rm min}.
\end{eqnarray}
In ~Eq.~(\ref{xi3}), we have denoted  the row vector 
$n_{\perp}^T ~=~\left(
n_{1\perp},\, n_{2\perp},\, n_{3\perp}\right).$

With the mean spin along the $z$ direction,
we have  $\hat{n}_{\perp} =\left(\cos\theta,\, \sin\theta,\, 0\right)$, and
the minimum value of the quadratic form
$(n_{\perp}^T\, T\, n_{\perp})_{\rm min}$ in ~Eq.~(\ref{xi3}) is fixed as
follows:
\begin{eqnarray}
\label{min}
(n_{\perp}^T\, T\, n_{\perp})_{\rm min}&=& \begin{array}{c} {\rm min}\\
\theta\end{array} \left(
t_{11}\,\cos^2\theta+t_{22}\,\sin^2\theta+t_{12}\sin\, 2\theta\right)\nonumber
\\
&=& \frac{1}{2}\left[(t_{11}+t_{22})-\sqrt{(t_{11}-t_{22}^2)+4\,
t_{12}^2}\right]\nonumber \\
&=& t_{\perp}^{(-)},
\end{eqnarray}
where $t_{\perp}^{(-)}$ is the least eigenvalue of $T_{\perp}$ (see
Eq.~(\ref{tperp})). We finally have
\begin{equation}
\label{xifinal}
\xi^2=\frac{4}{N}\, (\Delta\, S_{\perp})^2_{\rm min}=\left(1+(N-1)\,
t_{\perp}^{(-)}\right).
\end{equation}
Following similar lines we can also show that
\begin{equation}
\label{maxfluct}
\frac{4}{N}\, (\Delta\, S_{\perp})^2_{\rm max}=\left(1+(N-1)\,
t_{\perp}^{(+)}\right),
\end{equation}
which relates the eigenvalue $t_{\perp}^{(+)}$ of $T_{\perp}$ to the maximum
collective fluctuation  $(\Delta\, S_{\perp})^2_{\rm max}$
 orthogonal to the mean spin direction.
Substituting Eqs.~(\ref{xifinal}), (\ref{maxfluct})  and expressing
$s_{0}=\frac{2}{N}\vert\langle\vec{S}\rangle\vert$ in Eq.~(\ref{ss2I5}), we get
\begin{equation}
\label{invsq}
{\cal I}_5= \frac{8\,
\vert\langle\vec{S}\rangle\vert^2}{\left(N(N-1)\right)^2}\,
\left(\xi^2-1\right)\,
(\frac{4}{N}\, (\Delta\, S_{\perp})^2_{\rm max}-1).
\end{equation}
Having related the local invariant ${\cal I}_5$ to collective spin observables,
we now proceed to show  that
 ${\cal I}_5<0$  {\em iff} $\xi^2<1$ i.e., {\it iff} the state is spin
squeezed.

It has been shown~\cite{Hor96} that positivity of any arbitrary two-qubit
density operator imposes the bound
 $-1\leq t_{11},\ t_{22},\ t_{33}\leq 1, $ on  the diagonal elements of the
correlation matrix $T$. This bound, together
with the unit trace condition ${\rm Tr}\, (T)=1$ on the correlation matrix of a
symmetric two-qubit state, leads to the
identification that {\em only one} of the diagonal elements of $T$ can be
negative. This in turn implies that\break
$t_{\perp}^{(+)}=\frac{1}{N-1}\,\left( \frac{4}{N}(\Delta\, S_{\perp})^2_{\rm
max}-1\right)\geq 0$.

Therefore, a symmetric multiqubit state is spin-squeezed ($\xi^2<1$)
{\em iff}\, the local invariant ${\cal I}_5<0$.$\Box$

Further, from the structure of the invariant ${\cal I}_5$ given in
Eq.~(\ref{s1s2s3}), it is clear that
one of the eigenvalues $t_1, \, t_{2}$ or $t_3$  of the correlation matrix $T$
must be negative in order that
${\cal I}_5<0$. Thus the invariant ${\cal I}_1=t_1\, t_2\, t_3$ is also
negative, when the system is spin squeezed.

We now explore other collective signatures of pairwise entanglement, which are
manifestations of {\em negative values} of the invariants   ${\cal I}_4$ and  ${\cal I}_4-{\cal 
I}_{3}^2$.

It is easy to see that the invariant ${\cal I}_4$ assumes a simple form, when
the average spin vector is oriented
along the $z$-axis:
\begin{equation}
\label{I4New}
{\cal I}_4= s^T\, T\, s=s_{0}^2\, t_{33}.
\end{equation}
So, ${\cal I}_4$ is negative {\em iff}\, $t_{33}<0$. Using Eqs.~(\ref{vars}) and
(\ref{vart}),
we express ${\cal I}_4$ in terms of collective observables as
\begin{equation}
\label{I42}
{\cal I}_{4}=\frac{4}{N^2\,(N-1)}\, \vert\langle\vec{S}\rangle \vert ^2\,
\left( \frac{4}{N}\langle (\vec{S}\cdot \hat{n})^2\rangle -1\right),
\end{equation}
where $\hat{n}$ denotes a unit vector along the direction of mean spin. We
therefore read
from Eq.~(\ref{I42}), that {\em the average  of the squared spin component,
along the
 mean spin direction, reduced below the value  $N/4$,  signifies pairwise
entanglement in symmetric $N$-qubit system.}

We note that ${\cal I}_5$ is not negative, when ${\cal I}_4\leq 0$. This is
because  $t_{33}\leq 0$ implies
 $t_{\perp}^{(\pm)}\geq 0$, as  $t_{\perp}^{(+)}+t_{\perp}^{(-)}+t_{33}=1,$
(unit trace condition) and \break $-1\leq t_{\perp}^{(\pm)},\
t_{33} \leq 1, $ (positivity condition~\cite{Hor96}). Therefore,
{\em spin squeezing and $\langle (\vec{S}\cdot
\hat{n})^2\rangle\leq \frac{N}{4}$ are two mutually exclusive
criteria of pairwise entanglement}. However, it is obvious from
the structure of the invariant ${\cal I}_{4}$, as given in
Eq.~(\ref{s1s2s3}), that ${\cal I}_4 \leq 0$ implies  ${\cal I}_1=t_1\, t_2\,
t_3\,\leq 0.$

Now we relate ${\cal I}_4-{\cal I}_3^2$ to collective variables:
\begin{widetext}
\begin{eqnarray}
\label{I4I3}
{\cal I}_4-{\cal I}_3^2&=& s_0^2\, (t_{33}-s_0^2)\nonumber \\
&=& \frac{4}{N^2}\, \vert \langle\vec{S}\rangle\vert^2\, \left(
\frac{4}{N(N-1)}\,
\langle(\vec{S}\cdot \hat{n})^2 \rangle -\frac{1}{(N-1)}-
\frac{4}{N^2}\,\vert\langle\vec{S}\rangle\vert^2\right)\nonumber \\
&=& \frac{16}{N^3(N-1)}\, \vert \langle\vec{S}\rangle\vert^2\,\left(
 \langle(\vec{S}\cdot \hat{n})^2\rangle- \frac{N}{4}\,-\frac{(N-1)}{N}\,
\vert\langle\vec{S}\rangle\vert^2\right).
\end{eqnarray}
\end{widetext}
\begin{table}
\caption{\label{tab:table1} Classification of pairwise entanglement 
 in terms of two-qubit local invariants.}  
\begin{tabular}{|c|c|c|}\hline
\multicolumn{2}{|c|}
{Criterion of pairwise entanglement}  & Collective 
behaviour to look for \\ \hline 
& &  \\
& ${\cal I}_5\leq 0$ & $(\Delta S_\perp)^2_{\rm min} \leq \frac{N}{4}$\\  
& & \\
\cline{2-3} 
& & \\
${\cal I}_3\neq 0$\ & ${\cal I}_4\leq 0$ &  
$\langle(\vec{S}\cdot\hat{n})^2\rangle \leq \frac{N}{4}$ \\
& & \\
 \cline{2-3}
& & \\
 & ${\cal I}_4 > 0,\  {\cal I}_4-{\cal I}_3^2 < 0$   
  & $ \frac{N}{4}< \langle(\vec{S}\cdot\hat{n})^2\rangle < 
\frac{N}{4}+\frac{(N-1)}{N}\, \vert\langle\vec{S}\rangle\vert^2$ \\ 
& & \\
\hline 
& & \\
 ${\cal I}_3 =0$ & ${\cal I}_1 < 0$  & $\langle S_i^2\rangle < \frac{N}{4}$\\   
& & for any direction $i=1,\,2,\, 3,$ so that\\
& &  $\langle (S_i\, S_j+S_j\, S_i)\rangle=0;\ {\rm for\ } i\neq j$ \\
& & \\
\hline
\end{tabular}
\label{tab1}
\end{table}

Negative value of the combination ~$~{\cal I}_4-{\cal I}_3^2~$~ manifests
itself through
$\langle(\vec{S}\cdot \hat{n})^2\rangle<  \frac{N}{4}+\frac{(N-1)}{N}\,
\vert\langle\vec{S}\rangle\vert^2 .$
From Eqs.~({\ref{I42}) and (\ref{I4I3}), we conclude that
{\em pairwise entanglement  resulting from
$ {\cal I}_3\neq 0,\ {\cal I}_{4}> 0$, but
${\cal I}_4-{\cal I}_3^2< 0,  $  is realised, whenever}
$$\frac{N}{4}< \langle(\vec{S}\cdot \hat{n})^2\rangle <
\frac{N}{4}+\frac{(N-1)}{N}\, \vert\langle\vec{S}\rangle\vert^2.$$

In the cases, where the qubits have no preferred orientation, i.e., when 
$\vert\langle\vec{S}\rangle\vert=0$,  all the local invariants
 except ${\cal I}_{1}$ and ${\cal I}_2$ are zero.  In such situations, pairwise
entanglement manifests itself ~\cite{footnote3}
\  ${\cal I}_1<0.$ In terms of collective observables, we have
\begin{equation}
{\cal I}_1=\left(\frac{4}{N(N-1)} \right)^3\, \displaystyle\prod_{i=1}^3\left(
\langle S_{i}^2\rangle-\frac{N}{4}\right).
\end{equation}
{\em Negative value of ${\cal I}_1$ shows up through $\langle
S_{i}^2\rangle<\frac{N}{4}$}  along the  axes $i=1,\, 2\, $ or 3,
which are fixed by verifying   $\langle (S_i S_j+S_j S_i)\rangle=0; \  i\neq
j$, as $T$ is diagonal with such a choice of the axes.

Finally, we arrive at a classification scheme, as depicted in Table.1, for
pairwise entanglement in symmetric multi-qubit states. Next, we proceed to relate our invariant 
criteria with the recently proposed generalized spin squeezing inequalities~\cite{cirac} for 
pairwise entanglement
\begin{eqnarray}
\label{ineq}
\frac{4\langle\Delta S_k\rangle^2}{N}<1-\frac{4\langle S_k^2\rangle}{N^2}
\end{eqnarray}
where $S_k=\vec{S}\cdot\hat{k}$\,; with $\hat{k}$ denoting an arbitrary unit vector. We consider 
different situations as discussed below:\\
(i) Let us now choose $\hat{k}=\hat{n}_\perp,$ a direction orthogonal to the mean spin vector 
$\langle\vec{S}\rangle$. The inequality given by
~Eq.~(\ref{ineq}) reduces to
$$(\Delta S_{n_\perp})^2 < \frac{N}{4}.$$
Minimizing the left hand side of this inequality gives $(\Delta S_\perp)_{min}^2 < \frac{N}{4}$ 
and corresponds to the condition ${\cal I}_5<0$ on the local invariant.\\
(ii) If $\hat{k}$ is aligned along the mean spin direction i.e.,
$\hat{k}=\hat{n}$ with $\hat{n}=\frac{\langle\vec{S}\rangle}{\vert \langle\vec{S}\rangle \vert},$ 
the generalized spin squeezing inequalities (see ~Eq.~(\ref{ineq})) reduce to the form,

\begin{eqnarray}
\label{rineq}
\langle(\vec{S}\cdot\hat{n})^2\rangle<\frac{N}{4}+\frac{(N-1)}{N}\vert \langle \vec{S}\rangle 
\vert^2.
\end{eqnarray}
Note that the condition ${\cal I}_4<0$ on the local invariant leads to the collective signature 
(see Table.1)
$$\langle(\vec{S}\cdot\hat{n})^2\rangle<\frac{N}{4},$$
which is a stronger restriction than that given by ~Eq.~(\ref{rineq}).
Further, if ${\cal I}_4<0$ but ${\cal I}_4-{\cal I}_3^2<0$ we obtain the inequality (see Table.1)
$$ \frac{N}{4}<\langle(\vec{S}\cdot\hat{n})^2\rangle<\frac{N}{4}+\frac{(N-1)}{N}\vert \langle 
\vec{S}\rangle \vert^2,$$
which covers the remaining range of possibilties contained in the
generalized spin squeezing inequalities of ~Eq.~(\ref{rineq}) with $\hat{n}$ along the mean spin 
direction.\\
(iii) If the average spin is zero for a given state, we have $\langle S_k\rangle=0$  for all 
directions $\hat{k}$. The inequalities of Korbicz {\it et al.}\cite{cirac} assume a simple form 
$\langle S_k\rangle^2<\frac{N}{4}$ in this case, which obviously corresponds to ${\cal I}_3=0$ 
and ${\cal I}_1<0.$

Now we consider specific examples of symmetric
multi-qubit states.
\subsection{Dicke state}
Dicke states,  $\vert S=\frac{N}{2},M \rangle\,  ;\ -S\leq M \leq S$,  are collective symmetric 
multi-atom  states, which were shown~\cite{Dic54}
to exhibit enhanced spontaneous emission rates (superradiance)
in  atom-field interactions.
For any random pair of atoms, drawn from a Dicke state, the state variables are
given by
\begin{eqnarray}
\label{dicst}
\vec{s}\equiv \left(0,\,0,\,\displaystyle{\frac{2M}{N}}\right),\hskip 2in \nonumber\\ 
T={\rm diag}\,\left(\frac{N^2-4M^2}{2N(N-1)},\, \frac{N^2-4\,M^2}{2N(N-1)},\,\frac{4M^2-N}{N(N 
-1)}\right).
\end{eqnarray}
The two-qubit local invariants associated with a Dicke state are
\begin{equation}
\label{dicinv}
\begin{array}{l}{\cal I}_1=\left(\frac{N^2-4M^2}{2N(N-1)}\right)^2\,
\left(\frac{4M^2-N}{N(N-1)}\right),  \\
   {\cal I}_2=2\,\left(\frac{N^2-4M^2}{2N(N-1)}\right)^2+\left(\frac{4M^2-N}{N(N-1)}
\right)^2   \\
{\cal I}_3=\frac{4M^2}{N^2}, \hskip 0.3in {\cal I}_4=
{\cal I}_3\, \left(\frac{4M^2-N}{N(N-1)} \right)\\
{\cal I}_5=8\, {\cal I}_3\,
\left(\frac{N^2-4M^2}{4\,N(N-1)}\right)^2, \hskip 0.3in  {\cal I}_6=0.
\end{array}
\end{equation}
 Further, the combination ${\cal I}_4-{\cal I}_3^2$ of invariants, is given by
 \begin{equation}
{\cal I}_4-{\cal I}_3^2= \left(\frac{4M^2-N^2}{N^2(N-1)}\right)
\,{\cal I}_3.
\end{equation}
We identify three different cases here:
$(i)\  M=\pm\frac{N}{2}, \ (ii)\ M=0, \ (iii)\ M\neq\pm\frac{N}{2},\,0 $
Case (i) corresponds to the situation in which all the qubits are {\em spin-up} ({\em 
spin-down}). The collective  state, corresponding to this case, is obviously  a product state. 
The invariants in this case are given by
\begin{equation}
{\cal I}_1={\cal I}_5=0,
\ \ \ \ {\cal I}_2={\cal I}_3={\cal I}_4=1,
\end{equation}
reflecting  that the  system is  not entangled.
In case (ii) the invariants ${\cal I}_3={\cal I}_4={\cal I}_5=0$, while the non-zero invariant
\begin{equation}
{\cal I}_1=-\frac{1}{4}\left(\frac{N}{N-1}\right)^3
\end{equation}
assumes negative value. So, the Dicke state $\left|\frac{N}{2},\, 0\right\rangle$, (with even 
number of atoms), exhibits pairwise entanglement.
In case (iii) the invariant	
${\cal I}_4$ is bound by $ -\frac{1}{N-1}<{\cal I}_4<1,$ and  the combination
  ${\cal I}_4-{\cal I}_3^2$ is always negative, thus revealing pairwise entanglement in  Dicke 
atoms in this case too.
\subsection{Kitagawa-Ueda state generated by one axis twisting Hamiltonian}		
Kitagawa and Ueda~\cite{Kit93} had proposed the generation of
correlated $N$-qubit states, which are spin squeezed,  through  the Hamiltonian evolution
\begin{equation}
\label{ku}
\left|\Psi_{\rm K-U}\right\rangle= e^{-iS_3^2\, \chi\, t}\,\left|S, -S\right\rangle
;\ \ S=\frac{N}{2},
\end{equation}
referred to  as {\em one-axis twisting mechanism}. The effective Hamiltonian
$H=S_{3}^2\, \chi$, has already been employed to produce entangled states of four 
particles~\cite{Sac00}. Collisional interactions between
atoms in two-component Bose-Einstein condensation are also modeled using this one-axis twisting 
Hamiltonian~\cite{Sor101}.

A random pair of qubits drawn from the Kitagawa-Ueda state are
characterized by the following parameters:
\begin{equation}
\label{kus}
\vec{s}=\left(0,\, 0,\, -\cos^{(N-1)}(\chi\, t)\right),
\end{equation}
and the correlation matrix elements given by
\begin{eqnarray}
t_{11}&=&t_{13}=t_{23}=0, \ \ \ \ t_{12}= \cos^{(N-2)}(\chi\, t)\, \sin(\chi\, t),\nonumber \\
t_{22}&=&\frac{1}{2}\, \left( 1-\cos^{(N-2)}(2\chi\, t)\right), \nonumber\\
 t_{33}&=& \frac{1}{2}\, \left( 1+\cos^{(N-2)}(2\chi\, t)\right).
\end{eqnarray}
We give below, the two-qubit invariants associated with the Kitagawa-Ueda  state:
\begin{eqnarray}
 \label{kuinv}
{\cal I}_1&=&-\frac{1}{2}\, \cos^{2(N-2)}(\chi\, t)\, \sin^2(\chi\, t)\,
\left( 1+\cos^{(N-2)}(2\chi\, t)\right), \nonumber \\
{\cal I}_2&=&2\, \cos^{2(N-2)}(\chi\, t)\, \sin^2(\chi\, t)+
\frac{1}{2}\,  \left( 1+\cos^{2(N-2)}(\chi\, t)\right), \nonumber  \\
{\cal I}_3&=&\cos^{2(N-1)}(\chi\, t), \ \ \ {\cal I}_4=\frac{1}{2}\, {\cal I}_3\,
\left( 1+\cos^{(N-2)}(2\chi\, t)\right), \nonumber  \\
{\cal I}_5&=&-2\, {\cal I}_3\, \cos^{2(N-2)}(\chi\, t)\, \sin^2(\chi\, t),\ \ {\cal I}_6=0.
\end{eqnarray}
\begin{figure}
 \includegraphics*[width=3.5in,keepaspectratio]{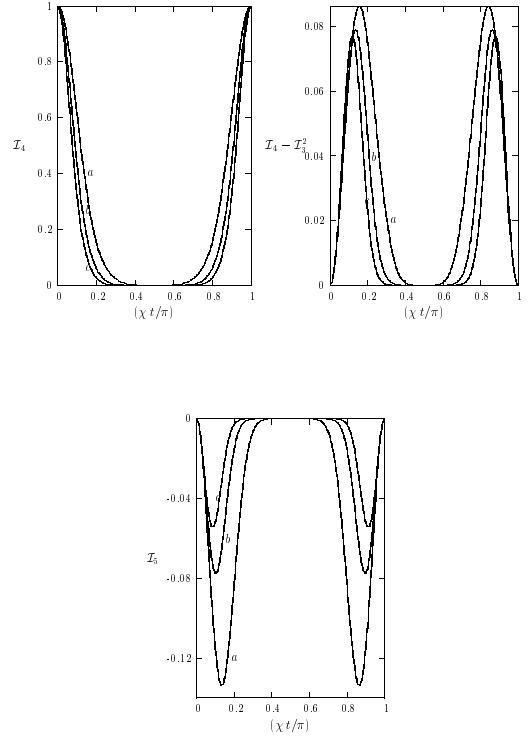}
  \caption{The invariants ${\cal I}_4,\ {\cal I}_4-{\cal I}_3^2 \  {\rm and}\ {\cal I}_5$,  
corresponding to a $N$-qubit Kitegawa-Ueda state. Curve $a:\, N=4,\, b: \, N=6, \ {\rm and}\  c: 
\, N=8.$}
  \label{fig:GeoSpace}
\end{figure}
We see that pairwise entanglement is manifest through  ${\cal I}_5<0$,
collective signature of which is spin squeezing. In Fig. 1, we have plotted ${\cal I}_4,\ {\cal 
I}_4-{\cal I}_3^2 \  {\rm and}\ {\cal I}_5$, for different
values of $N$.
\begin{figure}
 \includegraphics*[width=3.5in,keepaspectratio]{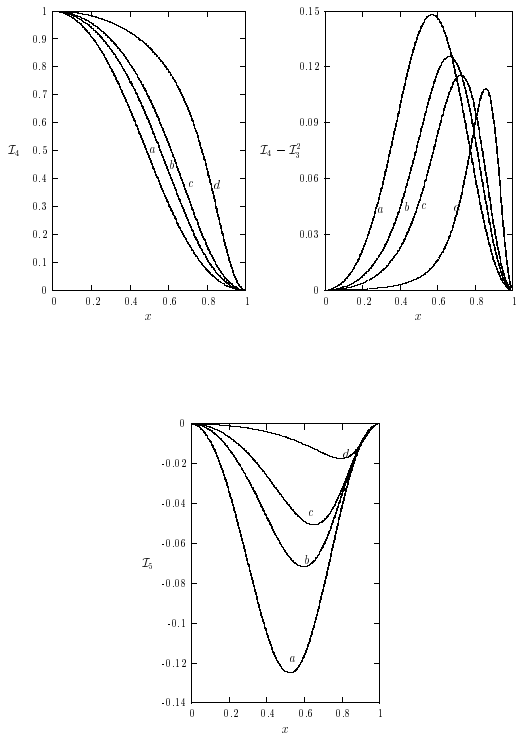}
  \caption{The invariants  ${\cal I}_4,\ {\cal I}_4-{\cal I}_3^2 \  {\rm and}\ {\cal I}_5$,  
associated with the steady state of $N$ two level atoms interacting with  squeezed radiation. 
Curve $a:\, N=4,\, b: \, N=6, \, c: N=8,\  {\rm and}\  d: \, N=20.$}
  \label{fig:GeoSpace}
\end{figure}
\subsection{Steady state of atoms irradiated by a squeezed bath}
Steady state, realized  in the interaction of a collection of $N$  two-level atoms with
squeezed radiation, has been shown~\cite{Agarwal90} to be a pure state (for even $N$) given by
\begin{equation}
\label{agarwal}
\left|\Psi_{0}\right\rangle=A_{0}\, {\rm exp}\, \left(\theta\, S_{3}\right)\,
{\rm exp}\left(-i\, \frac{\pi}{2}\, S_2\right)\, \left| S=\frac{N}{2},\, 0\right\rangle ,
\end{equation}
for certain values of external field strength and detuning parameters. In Eq.~(\ref{agarwal}),
$\theta=\frac{1}{2}\, \ln \left(\tanh (2|\xi|)\right);\ \xi$ is the squeezing parameter
associated with the input radiation, with mean photon number $\bar{n}=\sinh^2|\xi|$ and
$A_0$ is the normalization constant.

Two-qubit state variables, corresponding to the collective atomic system, are given by
\begin{eqnarray}
\label{agarwalst}
\vec{s}&=&\left(0,\, 0, \, \frac{2\,\langle S_3\rangle}{N}\right),\ \ T= {\rm diag}\, (t_1,\, 
t_2,\, t_3), \nonumber \\
t_1&=& \frac{-2\, \langle S_3\rangle\, e^{-2|\xi|}-N }{N(N-1)},\ \
t_2= \frac{-2\, \langle S_3\rangle\, e^{2|\xi|}-N}{N(N-1)}\nonumber \\
t_3&=& \frac{4\, \langle S_3\rangle\, \cosh(2|\xi|)+N^2+N}{N(N-1)},
\end{eqnarray}
with
\begin{equation}
\label{agarwals3}
\left\langle S_3\right\rangle= \frac{\displaystyle\sum_{M=-S}^{S}\, M\, e^{2\, M\, \theta}\, 
\left(d^{S}_{M\, 0}(\frac{\pi}{2})\right)^2}
{\displaystyle\sum_{M=-S}^{S}\,  e^{2\, M\, \theta}\, \left(d^{S}_{M\, 
0}(\frac{\pi}{2})\right)^2}
\end{equation}
and the coefficient $d^{S}_{M\, 0}(\frac{\pi}{2})$ is given  by~\cite{Brink}
\begin{eqnarray}
d^{S}_{M\, 0}\left(\frac{\pi}{2}\right)&=&\frac{S!\, \sqrt{(S+M)!\, (S-M)!}}{2^S}\, \nonumber\\
& &\displaystyle\sum_{p=M}^{S-M}\,\frac{(-1)^p}{ (S-M)!\, p!\, (p-M)!\,  (S+M-p)!}.\nonumber \\
& &
\end{eqnarray}
The corresponding local invariants are listed below:
\begin{eqnarray}
\label{agarwalinv}
{\cal I}_1&=&t_1\,t_2\,t_3,\ \  {\cal I}_2=t_1^2+t_2^2+t_3^2 \nonumber \\
{\cal I}_3&=&\frac{4\, \left\langle S_3\right\rangle^2}{N^2}, \ \
{\cal I}_4= {\cal I}_3\, \frac{4\, \langle S_3\rangle\, \cosh(2|\xi|)+N^2+N}{N(N-1)} \nonumber \\
{\cal I}_5&=&\frac{{2\,\cal I}_3}{N^2\, (N-1)^2}\, \left(2\, \langle S_3\rangle\, 
e^{-2|\xi|}+N\right)\,
\left(2\, \langle S_3\rangle\, e^{2|\xi|}+N\right),\nonumber\\ {\cal I}_6&=&0.
\end{eqnarray}
In Fig. 2 we have plotted the invariants ${\cal I}_4,\ {\cal I}_4-{\cal I}_3^2 \  {\rm and}\ 
{\cal I}_5$,   as a function of the parameter $x=e^{2\,\theta}=\sqrt{\frac{\bar{n}}{\bar{n}+1}}$, 
for different values of $N$. These  plots demonstrate  that the invariant ${\cal I}_5$ is 
negative, highlighting the pairwise entanglement (spin squeezing) of the
atomic state.\\

\section{\label{sec:level3}Summary}
In summary, we have shown that a set of six local invariants $\{ {\cal I}_1\ -\  {\cal I}_6\}$, 
associated
with the two-qubit partition of a symmetric multiqubit system, completely characterizes  the 
pairwise entanglement
properties of the collective state. For symmetric separable states, we have  proved that the 
entanglement invariants
${\cal I}_1,\, {\cal I}_4,\, {\cal I}_5$ and ${\cal I}_4-{\cal I}_3^2$ assume {\em non-negative} 
values.
 We have proposed a detailed classification scheme,  for pairwise entanglement in symmetric
multiqubit system, based on  {\em negative} values of the  invariants
${\cal I}_1,\, {\cal I}_4,\, {\cal I}_5$ and ${\cal I}_4-{\cal I}_3^2$ .  Our scheme also relates 
appropriate
collective {\em non-classical} features, which  can be
identified in each case of pairwise entanglement. Specifically, we have shown, collective spin 
squeezing in symmetric multi-qubit states is a manifestation of ${\cal I}_5<0$. Moreover,
we have related our criteria, which are essentially given in terms of invariants of the quantum 
state, to the recently proposed generalized spin squeezing inequalities for bipartite 
entanglement. We have studied (i) Dicke state, (ii) Kitagawa-Ueda state
generated by the the application of one-axis twisting Hamiltonian, and (iii) the steady state of 
a collection  of two-level atoms interacting with  squeezed radiation, making use of the proposed  
characterization. We hope that our analysis
may  provide useful insights in understanding higher order quantum correlations in multi-particle 
states.


\begin{thebibliography}{1}
\bibitem{footnote1} The qubit labels 1 and 2  in
$\langle(\sigma_{1\,i}\sigma_{2\, j})\rangle$ are only instructive. Infact, the
average values of
two qubit correlations are identical for any random pair of qubits belonging to
a
symmetric $N$-qubit system.
\bibitem{footnote3}
From the  unit trace condition $t_1+t_2+t_3=1$ and the positivity condition
$-1\leq t_1,\ t_2, t_{3}\leq 1,$ it is clear that
$\det T <0$ also leads to ${\rm Tr}\,(T^2)>1$ i.e., ${\cal I}_1<0$ and ${\cal
I}_2 >1$ both imply pairwise entanglement.
\end{thebibliography}
\end{document}